&pdflatex

\documentclass[12pt]{iopart}
\usepackage{iopams}
\usepackage{setstack}
\usepackage{graphicx}

\newcommand{\cA}{\ensuremath{\mathcal{A}}}
\newcommand{\cS}{\ensuremath{\mathcal{S}}}

\newcommand{\cP}{\ensuremath{\mathcal{P}}}
\newcommand{\cT}{\ensuremath{\mathcal{T}}}
\newcommand{\cL}{\ensuremath{\mathcal{L}}}

\newcommand{\veps}{\varepsilon}

\begin{document}

\rightline{preprint KCL-PH-TH-2012-22}

\title[$\cP\cT$-Symmetric Interpretation of Double-Scaling]
{$\cP\cT$-Symmetric Interpretation of Double-Scaling}

\author[Bender, Moshe, and Sarkar]{Carl~M~Bender$^{a,}$\footnote{
{\footnotesize{\tt email: cmb@wustl.edu}}}, Moshe Moshe$^{b,}$\footnote{
{\footnotesize{\tt email: moshe@technion.ac.il}}}, Sarben
Sarkar$^{c,}$\footnote{{\footnotesize{\tt email: sarben.sarkar@kcl.ac.uk}}}}

\address{$^a$Department of Physics, Washington University, St. Louis MO 63130,
USA\\ $^b$Department of Physics, Technion -- Israel Institute of Technology,
32000 Haifa, Israel\\ $^c$Department of Physics, King's college London, Strand,
London WC2R 2LS, UK}

\date{today}

\begin{abstract}
The conventional double-scaling limit of an O($N$)-symmetric quartic quantum
field theory is inconsistent because the critical coupling constant is negative.
Thus, at the critical coupling the Lagrangian defines a quantum theory with an
upside-down potential whose energy appears to be unbounded below. Worse yet, the
integral representation of the partition function of the theory does not exist.
It is shown that one can avoid these difficulties if one replaces the original
theory by its $\cP\cT$-symmetric analog. For a zero-dimensional O($N$)-symmetric
quartic vector model the partition function of the $\cP\cT$-symmetric analog is
calculated explicitly in the double-scaling limit.
\end{abstract}


\submitto{\JPA}


\section{Introduction}
\label{s1}
The techniques of $\cP\cT$ quantum mechanics have been used to solve several
long-standing problems, such as the violation of unitarity in the Lee model
\cite{R1} and the appearance of ghosts in the Pais-Uhlenbeck model \cite{R2} and
in other field-theory models \cite{R3}. In this paper we clarify a serious
problem with the double-scaling limit in quantum field theory that in our
opinion has not yet been satisfactorily addressed; namely, that the critical
coupling constant $g_{\rm crit}$ is {\it negative}. Since $g_{\rm crit}<0$, the
double-scaling limit appears to be unphysical because near $g_{\rm crit}$ the
potential is upside-down and the energy appears to be unbounded below. However,
we show that if we approach the critical theory in a $\cP\cT$-symmetric
fashion, the resulting correlated limit gives a physically acceptable quantum
theory. Here, we study the double-scaling limit of a zero-dimensional
O($N$)-symmetric quartic vector field-theoretic model. (In a more detailed paper
we will show that the same approach works for theories in higher dimension
\cite{R4}.)

To explain the mathematical problem addressed in this paper we consider the
partition function $Z(g)$ for a toy zero-dimensional quartic quantum field
theory model:
\begin{equation}
Z(g)=\frac{1}{\sqrt{2\pi}}\int_{-\infty}^\infty dx\,e^{-x^2/2-gx^4/4},
\label{e1}
\end{equation}
where the coupling constant $g$ is assumed at first to be positive. The
weak-coupling series for $Z(g)$ is obtained by expanding the integrand of
(\ref{e1}) in powers of $g$ and then integrating term by term. (This is a
trivial application of Watson's Lemma \cite{R5}.) The result is the (divergent)
asymptotic-series representation for $Z(g)$:
\begin{equation}
Z(g)\sim\sum_{n=0}^\infty(-g)^n\frac{(4n-1)!!}{4^n n!}\qquad\left(g\to0^+
\right).
\label{e2}
\end{equation}
This series alternates in sign when $g>0$ and it can be Borel summed to recover
the integral representation (\ref{e1}).

The integral representation (\ref{e1}) for $Z(g)$ ceases to exist when $g$ is
negative. Thus, to study the behavior of $Z(g)$ for negative $g$ we evaluate the
integral for positive $g$ in terms of a parabolic cylinder function \cite{R6}:
\begin{equation}
Z(g)=(2g)^{-1/4}e^{1/(8g)}{\rm D}_{-1/2}\left(1/\sqrt{2g}\right).
\label{e3}
\end{equation}
We conclude from this formula that $Z(g)$ is a multiple-valued function of $g$;
it is defined on a four-sheeted Riemann surface and is complex when $g$ is
negative. Thus, to evaluate $Z(g)$ for $g<0$ we must define precisely the path
from positive to negative $g$. We obtain four possible values for $Z(g)$
depending on how we rotate in the complex-$g$ plane from $g>0$ to $g<0$.

Even though $Z(g)$ is a four-valued function of $g$, each term in the asymptotic
series (\ref{e2}) is single valued. The resolution of this apparent discrepancy 
lies in identifying the wedge-shaped region in which a series is asymptotic to
the function that it represents; this region is called a {\it Stokes wedge}.
(For a detailed discussion of Stokes wedges see Refs.~\cite{R5,R7}.) The angular
opening of the Stokes wedge for ${\rm D}_\nu(x)$ for $x>>1$ is $|{\rm arg}\,x|<
3\pi/4$ \cite{R5}. Thus, in the complex-$g$ plane the Stokes wedge is centered
about the real-$g$ axis and the full opening angle is $3\pi$. This means that 
the series (\ref{e2}) is asymptotic to $Z(g)$ in (\ref{e3}) on the negative-$g$
axes ${\rm arg}\,g=\pm\pi$ as $|g|\to0$. Thus, we have the paradoxical result
that on the negative axes (\ref{e2}) is real while $Z(g)$ is complex. This
happens because the imaginary part of $Z(g)$ is {\it subdominant} (exponentially
small); it is of order $e^{1/g}$, as can be determined by a saddle-point
calculation.

In this paper we are interested in the partition function for negative coupling
constant \cite {R8}. However, we will not be interested in the complex partition
function that one obtains by rotating $x^2+gx^4$ from positive $g$ to negative
$g$ in the complex-$g$ plane. Rather, we will keep the coupling constant $g$
fixed and perform the $\cP\cT$-symmetric limit of $x^2+gx^2(ix)^\veps$ as
$\veps$ goes from $0$ to $2$ \cite{R9}. The resulting partition function is
\begin{equation}
Z(g)=\frac{1}{\sqrt{2\pi}}\int_C dx\,e^{-x^2/2+gx^4/4},
\label{e4}
\end{equation}
where $C$ is a comlex contour of integration. The original integration path lies
on the real axis when $\veps=0$. The path rotates downward into the complex
plane as $\veps$ increases. When $\veps=2$, the final contour $C$ comes inward
from $x=\infty$ in the $45^\circ$ wedge $-7\pi/8<{\rm arg}\,x<-5\pi/8$ and goes
back out to $x=\infty$ in the $45^\circ$ wedge $-3\pi/8<{\rm arg}\,x<-\pi/8$. As
a consequence, this partition function is {\it real}. This is the
$N$-dimensional version of the $\cP\cT$-symmetric theory that is studied in
this paper.

This paper is organized as follows. Correlated limits are explained in
Sec.~\ref{s2} using several illustrative examples. Then the uncorrelated and
correlated expansions of an O($N$)-symmetric vector theory in zero dimensions
are investigated in Secs.~\ref{s3} and \ref{s4} and problems with implementing
the double-scaling limit are described. To solve these problems the vector
theory is reformulated as a $\cP\cT$-symmetric model in Sec.~\ref{s5}, and the
uncorrelated and correlated expansions of this $\cP\cT$-symmetric model are
derived in Secs.~\ref{s6} and \ref{s7}. Brief concluding remarks are given
in Sec.~\ref{s8}.

\section{Correlated limits}
\label{s2}
Correlated limits arise frequently in physical problems when there are two
parameters, say $\veps$ and $\alpha$, and $\veps$ is treated as small ($\veps
\ll1$) so that it plays the role of a perturbation parameter. If we treat
$\alpha$ as fixed, the solution $\cS(\veps,\alpha)$ to the problem is a formal
perturbation series in powers of $\veps$: $\cS(\veps,\alpha)\sim\sum_{n=0}^
\infty a_n(\alpha)\veps^n$. Usually, the perturbation series is a {\it
divergent} asymptotic series in which each term in the series is negligible
relative to the previous term as $\veps$ tends to $0$; that is, $a_n(\alpha)
\veps^n\ll a_{n-1}(\alpha) \veps^{n-1}$ as $\veps\to0$ for all $n$. A correlated
limit occurs when we do {\it not} treat $\alpha$ as fixed, but rather allow it
to tend to a limit as $\veps\to0$; that is, we take $\alpha$ to depend on
$\veps$: $\alpha=\alpha(\veps)$.

A nontrivial correlated limit arises if we choose the functional dependence so
that {\it all} terms in the perturbation series become comparable as $\veps\to
0$. When this happens, the series undergoes a transmutation in which it depends
on just {\it one parameter}, which we call $\gamma$. In this correlated limit
the perturbation series still diverges, but we sum the series for $\cS(\gamma)$
by using Borel summation. Correlated limits are remarkable in that $\cS(\gamma)$
is a {\it universal} function that reveals the essential features of the problem
while being insensitive to specific details. Often, $\cS(\gamma)$ is {\it
entire} (analytic for all $\gamma$).

This paper examines the correlated limit of ${\rm O}(N)$-symmetric quantum
field theories. Such theories have been used to model a variety of physical
phenomena because the large-$N$ expansion in powers of $1/N$ often reveals the
phase structure of the theory; quantities such as masses and Green's functions
can be expressed as asymptotic series having the form $\sum_{k=0}^\infty a_k
N^{-k}$ \cite{R10}. In the {\it double-scaling} limit $N\to\infty$ and $g\to
g_{\rm crit}$ in a correlated fashion in which all terms in the $1/N$ expansion
are of comparable size. In the correlated limit the sum of the series is no
longer dominated by early terms; the $N^{-k}$ power in the $k$th term is
compensated by a sizable coefficient $a_k$ \cite{R11}. Furthermore, this limit
is characterized by a {\it universal} function of the parameters that describes
the correlated limit \cite{R11}. In the correlated limit O($N$)-symmetric vector
models represent discretized branched polymers. Additionally, dynamically
triangulated random surfaces summed on different topologies can be represented
by matrix models in the double-scaling limit \cite{R7,R12}. 

We have devised three elementary examples to illustrate correlated limits:

{\bf Example~1:} {\it Behavior of a nonuniformly convergent Fourier sine series
near the boundary of its interval of convergence.} Here, the number of terms $N$
in the partial sum of the Fourier series is correlated with the distance $x$ to
the boundary. This limit is described by the Gibbs function $G(\gamma)={\rm Si}
(2\gamma)$ (the sine-integral function), where $N\to\infty$, $x\to0$, and
$\gamma\equiv Nx$. The Gibbs function is entire and it is {\it universal}
because it describes the behavior of the Fourier sine series for {\it any}
differentiable function $f(x)$ such that $f(0)\neq0$ and/or $f(\pi)\neq0$. To
explain the famous {\it universal} 18\% overshoot exhibited by all nonuniformly
convergent Fourier series at the boundary, we simply verify that $G'(\pi/2)=0$
and that $G(\pi/2)=1.18\ldots$.

{\bf Example~2:} {\it Laplace's method for the asymptotic expansion of
integrals.} To find the large-$N$ behavior of the Laplace integral
\begin{eqnarray}
Z(N)=\int_0^{\infty}dr\,e^{-NS(r)},
\label{e5}
\end{eqnarray}
we assume that $S'(r)>0$ for all $r\geq0$ and use repeated integration by parts
to obtain the complete asymptotic expansion of $Z(N)$ as $N\to\infty$ \cite{R5}:
\begin{eqnarray}
Z(N)\sim e^{-NS(0)}\sum_{k=1}^{\infty}N^{-k}\left[\frac{1}{S'(r)}\frac{d}{dr}
\right]^{k-1}\frac{1}{S'(r)}\bigg|_{r=0}.
\label{e6}
\end{eqnarray}
This is an uncorrelated large-$N$ expansion.

Laplace's method emerges as a correlated limit of integration by parts: Suppose
now that $S'(0)$ is small [but that the higher derivatives of $S(r)$ are not
small at $r=0$]. As $S'(0)\to0$, the $k$th term in (\ref{e6}) is approximated by
\begin{eqnarray}
N^{-k}[-2S''(0)]^{k-1}[S'(0)]^{1-2k}\Gamma(k-1/2)/\Gamma(1/2)
\label{e7}
\end{eqnarray}
because this has the greatest number of powers of $S'(0)$ in the denominator.
Consider the correlated limit $N\to\infty$, $S'(0)\to0$, where $\gamma^2\equiv
N[S'(0)]^2/S''(0)$ is a fixed parameter. [We assume that $S''(0)>0$ so that
$\gamma^2>0$.] In this limit (\ref{e6}) becomes
\begin{eqnarray}
Z(\gamma)\sim\frac{e^{-NS(0)}}{\sqrt{NS''(0)}}\sum_{k=0}^{\infty}(-2)^k
\gamma^{-2k-1}\frac{\Gamma(k+1/2)}{\Gamma(1/2)}.
\label{e8}
\end{eqnarray}
This series diverges, but we can obtain its Borel sum in terms of the parabolic
cylinder function ${\rm D}_{\nu}(z)$ \cite{R6}:
\begin{eqnarray}
Z(\gamma)\sim e^{-NS(0)}\exp\left(\gamma^2/4\right){\rm D}_{-1}(\gamma)/
\sqrt{NS''(0)}.
\label{e9}
\end{eqnarray}
The function $Z(\gamma)$ is entire. Also, it is universal because it depends
only on the two numbers $S(0)$ and $S''(0)$, and thus it applies {\it
universally} to all functions $S(r)$ with these particular values. [In contrast,
the uncorrelated series (\ref{e6}) depends on {\it all} derivatives of $S(r)$ at
$r=0$. ]

For the special value $\gamma=0$, ${\rm D}_{-1}(0)=\sqrt{\pi/2}$ gives the
famous result
\begin{eqnarray}
Z(N)\sim e^{-NS(0)}\sqrt{\pi/[2NS''(0)]} \quad(N\to\infty)
\label{e10}
\end{eqnarray}
of Laplace's method applied to (\ref{e5}). The asymptotic formula (\ref{e10}) is
a limiting case of the correlated limit (\ref{e9}) for which $S'(0)=0$ and $S''(
0)>0$. Thus, (\ref{e9}) describes the approach of $Z(\gamma)$ to Laplace's
asymptotic formula (\ref{e10}). To be precise, the asymptotic expansion of the
Laplace integral is controlled by the function $S(r)$ at its Laplace point (that
is, its maximum point). Laplace's formula (\ref{e10}) gives the asymptotic
behavior of the integral for the special case in which $S(r)$ is {\it level}
(has a vanishing derivative) at its Laplace point. Thus, the correlated limit
describes in a smooth and {\it universal} fashion what happens as the derivative
of $S(r)$ approaches 0 at the Laplace point, just as the Gibbs function
describes in a smooth and universal fashion how a nonuniformly convergent
Fourier series for $f(x)$ behaves as $x$ approaches the boundary of the
interval.

{\bf Example~3:} {\it Transition in a quantum-mechanical wave function between a
classically allowed region and a classically forbidden region.} This transition
is described by the universal Airy function ${\rm Ai}(\gamma)$, which is
obtained by summing the divergent WKB series in the correlated limit $\hbar\to
0$, $x\to0$ (where $x$ is the distance to the turning point), with the ratio
$\gamma=x^{3/2}/\hbar$ held fixed. Consider the one-turning-point problem for
the Schr\"odinger equation $\hbar^2\phi''(x)=Q(x)\phi(x)$, where $Q(x)$ is
linear in $x$ near the turning point at $x=0$: $Q(x)\sim ax$ ($x\to0$). When
$x\neq0$ the WKB approximation to $\phi(x)$ is
\begin{equation}
\phi_{\rm WKB}(x)=\exp\bigg[\frac{1}{\hbar}\int_0^x ds\sum_{n=0}^\infty\hbar^n
S_n(s)\bigg]~~(\hbar\to0),
\label{e11}
\end{equation}
where $S_0(x)=\pm\sqrt{Q(x)}$, $S_1(x)=-Q'(x)/[4Q(x)]$, and $S_n(x)$ obeys the
recursion relation $2S_0(x)S_n(x)=-S_{n-1}'(x)-\sum_{j=1}^{n-1}S_j(x)S_{n-j}(x)$
for $n\geq2$.

The WKB approximation $\phi_{\rm WKB}(x)$ is invalid at the turning point
because it blows up at $x=0$ while the exact solution $\phi(x)$ remains finite.
To see how the WKB solution behaves as $x$ approaches $0$, we choose the
negative solution for $S_0$: $S_0(x)\sim-\sqrt{ax}$. Then, $S_1(x)\sim-1/(4x)$
and the functions $S_n(x)$ have the asymptotic form
\begin{equation}
S_n(x)\sim-4^{-n}a^{1/2-n/2}x^{1/2-3n/2}s_n\quad(x\to0),
\label{e12}
\end{equation}
where $s_n=(4-3n)s_{n-1}-\frac{1}{2}\sum_{j=1}^{n-1}s_j s_{n-j}$ for $n\geq2$
and $s_0=s_1=1$. The numerical coefficients $s_n$ become complicated as $n$
increases, but an easy way to understand the WKB series is to exponentiate it:
\begin{equation}
\exp\left[\sum_{n=2}^\infty\veps^{n-1}\int dx\,S_n(x)\right]
=\sum_{k=0}^\infty (-1)^kg_k\gamma^{-k},
\label{e13}
\end{equation}
where $\gamma=a^{1/2}x^{3/2}/\hbar$. The numerical coefficients $g_n$ in the
exponentiated series (\ref{e13}) now have the simple form $g_k=\pi^{-1/2}\Gamma(
3k+1/2)9^{-k}/\Gamma(2k+1)$. The series on the right side of (\ref{e13}) is just
the asymptotic expansion of the Airy function ${\rm Ai}(z)$ as $z\to+\infty$:
\begin{equation}
{\rm Ai}(z)\sim\frac{z^{-1/4}}{2\sqrt{\pi}}e^{-\frac{2}{3}z^{3/2}}\sum_{k=0
}^\infty(-1)^k g_kz^{-3k/2}.
\label{e14}
\end{equation}
This series diverges but it is Borel summable; in the correlated limit $\hbar\to
0$, $x\to0$, $\gamma$ fixed, the solution to the Schr\"odinger equation is $\phi
(\gamma)=c{\rm Ai}(\gamma)$, where $c$ is an arbitrary constant. This shows that
the solution to the famous one-turning-point problem is a correlated limit. The
Airy function is an {\it entire} function of $\gamma$ and it is universal
because it is valid for all potentials $Q(x)$ that vanish linearly at the
turning point.

\section{Uncorrelated large-$N$ series for quantum field theory in zero
dimensions}
\label{s3}
The partition function
\begin{eqnarray}
Z=\int d^{N+1}x\,{\rm
exp}\left[-\frac{1}{2}\sum_{n=1}^{N+1}x_n^2-\frac{\lambda}{4}\left(\sum_{n=1}^{
N+1}x_n^2\right)^2\right]
\label{e15}
\end{eqnarray}
represents a zero-dimensional quartic quantum field theory having ${\rm O}(N+1)$
symmetry. The coupling constant $\lambda$ is assumed to be {\it positive} so
that the integral converges. To derive an uncorrelated large-$N$ expansion of
$Z$, we exploit the rotational symmetry by introducing the radial variable $r$,
$\sum_{n=1}^{N+1}x_n^2=Nr^2$, and we let $\lambda=g/N$. The partition function
now takes the one-dimensional form
\begin{eqnarray}
Z=\cA_{N+1}\int_0^\infty dr\,e^{-NL(r)},
\label{e16}
\end{eqnarray}
where $\cA_N=2\pi^{N/2}/\Gamma(N/2)$ is the surface area of an $N$-dimensional
sphere of radius $1$ and $L(r)=r^2/2+gr^4/4-\log r$. We emphasize that we must
assume that $g$ is positive so that the integral representation (\ref{e16}) for
the partition function $Z$ converges.

The integral (\ref{e16}) is a Laplace integral and the correlated expansion of
the Laplace integral (\ref{e5}) was considered earlier. However, in contrast
with $Z(N)$ in (\ref{e5}), $L'(r)$ is not positive. We will see that as a
result, the correlated limit of (\ref{e16}) lies in a different universality
class and is characterized by a different universal function; specifically, the
universal function for (\ref{e16}) is an Airy function rather than a parabolic
cylinder function.

The standard procedure (Laplace's method) for finding the large-$N$ asymptotic
behavior of the integral (\ref{e16}) begins by locating the Laplace points,
which are the zeros of $L'(s)=r+gr^3-1/r$. Just one Laplace point $r_0=\sqrt{(G-
1)/(2g)}$, where $G\equiv\sqrt {1+4g}$, lies in the range of integration $0\leq
r<\infty$. Laplace's method relies on the crucial fact that if $r_0$ is a {\it
global minimum} of $L(r)$, then as $N\to\infty$ the entire asymptotic expansion
of the integral is determined by the behavior of $L(r)$ on the infinitesimal
region $r_0-\veps<r<r_0+\veps$ containing $r_0$. [The Laplace point $r_0$ is a
global minimum because $L''(r)=1+3gr^2+1/r^2$ and $L''\left(r_0\right)=2G>0$.]
We thus Taylor expand $L(r)$ about $r_0$ and conclude that for large $N$,
$$Z\sim \cA_{N+1}e^{-NL\left(r_0\right)}\int_{r_0-\veps}^{r_0+\veps}dr\,
e^{-NG\left(r-r_0\right)^2}\exp\left[-\sum_{k=3}^\infty\frac{N}{k!}L^{(k)}
\left(r_0\right)\left(r-r_0\right)^k\right].$$

To evaluate this integral we translate the integration region by substituting
$t=r-r_0$ and then we perform the scaling $t=u/\sqrt{NG}$. Because the sum in
square brackets begins at $k=3$, all terms in the sum are small as $N\to\infty$.
Thus, we can expand the exponential in powers of $u$. Making only
transcendentally small errors, we extend the range of integration to $-\infty<
u<\infty$, and perform the Gaussian integrals term-by-term. This yields an
infinite series in powers of $1/N$:
\begin{equation}
Z\sim\frac{\cA_{N+1}e^{-NL\left(r_0\right)}}{\sqrt{NG/\pi}}\sum_{k=0}^\infty a_k
N^{-k}\quad(N\to\infty).
\label{e17}
\end{equation}
This is the full asymptotic expansion of $Z$ and it is the {\it uncorrelated}
large-$N$ expansion of the partition function (\ref{e15}).
Only integer powers of $N^{-1}$ appear in this expansion; there are no
half-odd-integer powers because Gaussian integrals over odd powers of $u$
vanish. The first three coefficients in this series are explicitly
$$a_0=1,\quad a_1=\frac{5-6G^2-G^3}{24G^3},\quad
a_2=\frac{385-924G^2-10G^3+684G^4+12G^5-143G^6}{1152G^6}.$$

\section{Correlated limit of the large-$N$ expansion (\ref{e17})}
\label{s4}
Proceeding
formally, we now attempt to construct a correlated large-$N$ limit of the
expansion in (\ref{e17}). For all terms in the expansion to have the same order
of magnitude, the correlated limit must be $N\to\infty$ and $g\to g_{\rm crit}=-
1/4$ (that is, $G\to0$) with $\gamma\equiv NG^3/2$. In this limit the asymptotic
approximation (\ref{e17}) becomes
\begin{equation}
Z\sim\frac{\cA_{N+1}e^{-NL\left(r_0\right)}}{\sqrt{NG/\pi}}\left(1+\frac{5}
{48\gamma}+\frac{385}{4608\gamma^2}+\ldots\right).
\label{e18}
\end{equation}

We recognize that the series in (\ref{e18}) is the asymptotic expansion for
large $\gamma$ of ${\rm Bi}(\gamma^{2/3})\sqrt{\pi}e^{-2\gamma/3}\gamma^{1/6}$.
[This is like the series in (\ref{e13}) for the Airy function ${\rm Ai}$, but it
does not alternate in sign.] Thus, we are tempted to conclude that the
correlated limit of the series (\ref{e17}) is
\begin{equation}
Z\sim\cA_{N+1}e^{NL(\sqrt{2})}2^{-1/6}\pi N^{-1/3}{\rm Bi}(\gamma^{2/3})e^{-2
\gamma/3}.
\label{e19}
\end{equation}
However, this correlated limit is invalid because it requires that $g<0$.
Furthermore, to obtain the Airy function ${\rm Bi}$ in (\ref{e19}), we have had
to sum a {\it nonalternating} divergent series; such a series is not Borel
summable.

\section{$\cP\cT$-symmetric reformulation of the theory}
\label{s5}
In place of (\ref{e15}) we consider the O($N+1$)-symmetric partition function
$Z={\rm Re}\int d^{N+1}x\,e^{-L}$, where we assume that $N$ is an {\it even}
integer. The Lagrangian $L$ has the form
\begin{equation}
L=\frac{1}{2}\sum_{j=1}^{N+1}x_j^2+\frac{\lambda i^\veps}
{2+\veps}\Bigg(\sum_{j=1}^{N+1}x_j^2\Bigg)^{1+\veps/2}.
\label{e20}
\end{equation}
The multiple integral above is taken on the real axis and it converges if
$\veps<1$.

We let $\lambda=gN^{-\veps/2}$ and again introduce the radial variable $r$ by
$\sum_{n=1}^{N+1}x_n^2=Nr^2$. The crucial assumption that $N$ is {\it even}
allows us to extend the radial integral to the entire real-$r$ axis:
\begin{eqnarray}
Z=\frac{1}{2}\cA_{N+1}\int_{-\infty}^\infty dr\,e^{-N\cL(r)},
\label{e21}
\end{eqnarray}
where $\cL=r^2/2+gr^2(ir)^\veps/(2+\veps)-\log r$. This Laplace integral is real
because the integrand is $\cP\cT$ symmetric; that is, the integrand is invariant
under $r\to-r$ and $i\to-i$ \cite{R9}. Without its logarithm term, $\cL$ has a
standard $\cP\cT$-symmetric structure that has been studied in great detail
\cite{R9}. The logarithm in the exponent does not make the integral complex. [We
can make $\cL$ explicitly $\cP\cT$ symmetric by including an additive constant
$\log r\to\log(ir)$ and taking the branch cut to lie on the negative-imaginary
axis. The additive constant $\log i$ has no effect on the forthcoming
steepest-descent analysis.] To achieve the $\cP\cT$-symmetric partition function
(\ref{e21}) we have had to work in a space of {\it odd} dimension $N+1$. This
requirement can be traced to the absence of a distinct parity operator in
even-dimensional space, where the sign of $\vec{x}$ can be changed by a
rotation.

To obtain a quartic theory as $\veps\to2$, we must redefine the boundary
conditions on the integral (\ref{e21}) accordingly: For any $\veps\geq0$ the
integral converges if the integration contour lies inside a pair of $\cP
\cT$-symmetric Stokes wedges centered about $-\pi\veps/(4+2\veps)$ and $-(4\pi+
\pi\veps)/(4+2\veps)$. The wedges have angular opening $\pi/(2+\veps)$ and
contain the real-$r$ axis if $\veps<1$. As $\veps$ increases above $1$, the
wedges rotate downward into the complex plane and become narrower. At $\veps=2$
the wedges are centered about $-\pi/4$ and $-3\pi/4$ and have angular opening
$\pi/4$.

\begin{figure}[t!]
\begin{center}
\includegraphics[scale=0.51]{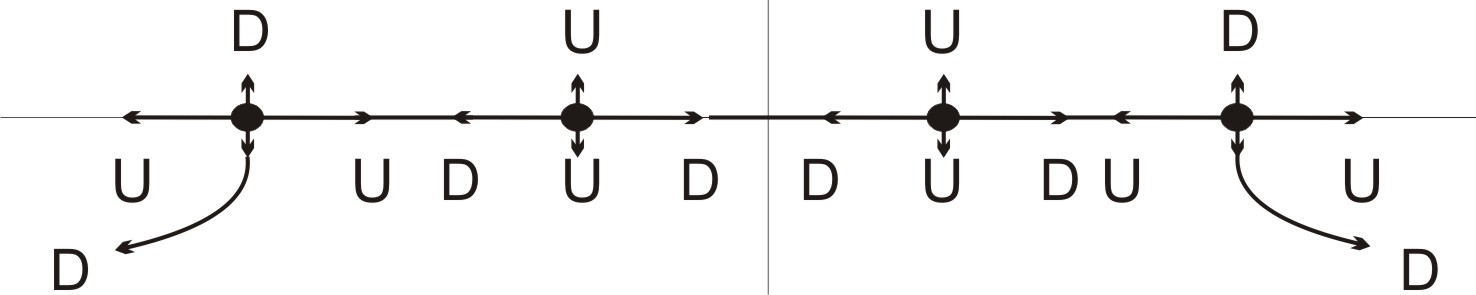}
\end{center}
\caption{The four saddle points [zeros of $\cL'(r)$] for the case $g<1/4$ with
the up and down directions indicated by the letters U and D. The
steepest-descent path is $\cP\cT$-symmetric (left-right symmetric) and runs
along the real axis between the distant pair of saddle points. Then it turns
downward, curves, and asymptotes at $-\pi/4$ and at $-3\pi/4$. For large $N$ the
contribution to the integral is localized at the pair of saddle points nearest
the origin.}
\label{F1}
\end{figure}

\section{Uncorrelated limit of the $\cP\cT$-symmetric theory (\ref{e21})}
\label{s6}
To find
the large-$N$ asymptotic behavior of $Z$ in (\ref{e21}) when $\veps=2$, we use
the {\it method of steepest descents} \cite{R5}. The saddle points are the zeros
of $\cL'(r)=r-gr^3-1/r$. There are four saddle points, which are the roots of
$r_0^2=(1\pm\sqrt{1-4g})/(2g)$. If $g<1/4$, the saddle points are all real and
are shown in Fig.~\ref{F1}. (If $g>1/4$, the saddle points are all imaginary,
but this does not affect the asymptotic analysis of the double-scaling limit.)

We determine the {\it directions} of the saddle points by calculating $\cL''(r)=
1-3gr^2+1/r^2$. At the saddle points $\cL''=\mp2\sqrt{1-4g}$. For the minus
(plus) sign the steepest-descent path moves away from the saddle point in the
imaginary (real) direction. Thus, the complete steepest-descent path, as shown
in Fig.~\ref{F1}, follows the real-$r$ axis until it reaches the distant pair of
saddle points. Then it turns downward and curves off at the angles $-\pi/4$ and
$-3\pi/4$. The entire contribution to the uncorrelated limit comes from an
infinitesimal region surrounding the two saddle points closest to $r=0$.

\section{Correlated limit of the $\cP\cT$-symmetric theory}
\label{s7}
\begin{figure}[h!]
\begin{center}
\includegraphics[scale=0.38]{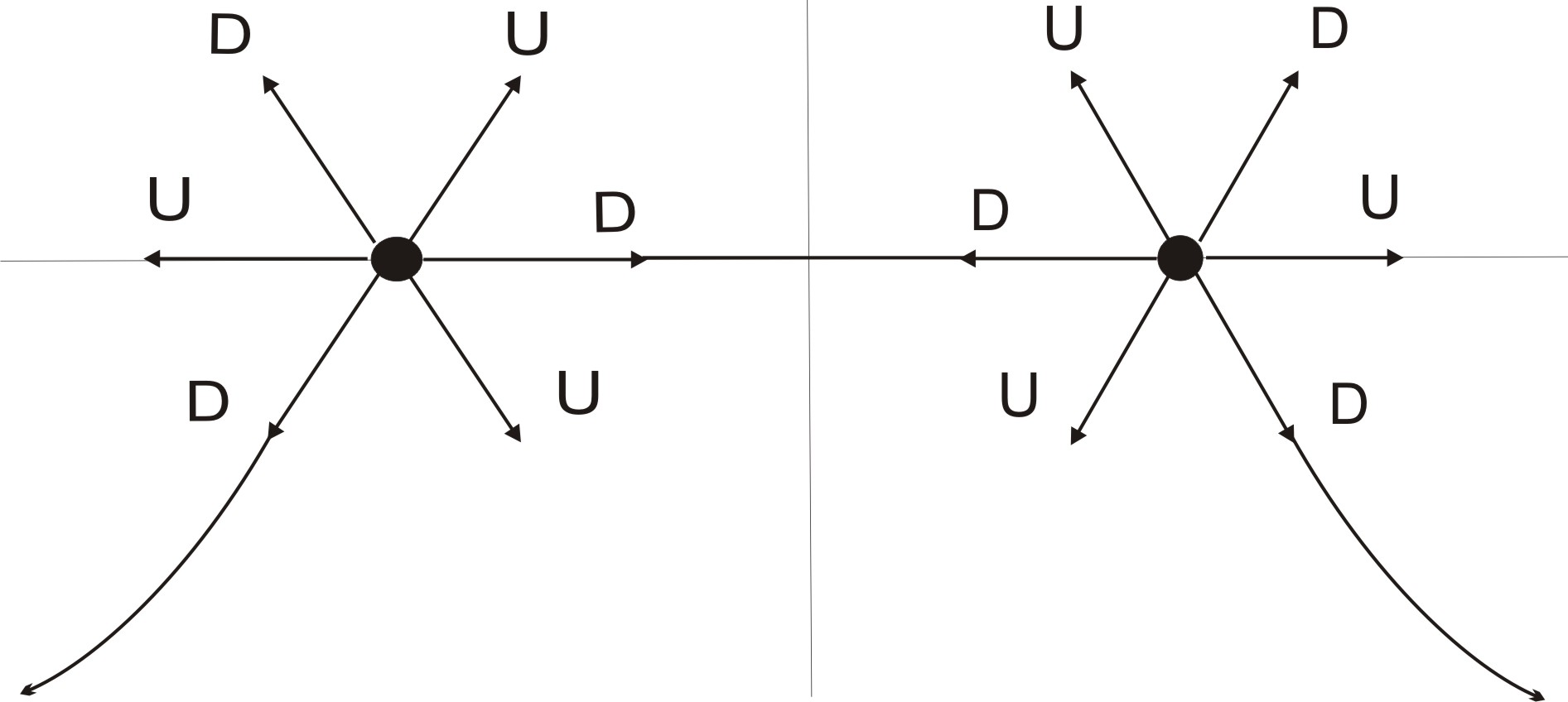}
\end{center}
\caption{Correlated limit of Fig.~\ref{F1}. At $g=g_{\rm crit}$ the quadratic
saddle points coalesce to form cubic saddle points with up and down directions
indicated by U and D. The steepest path now runs along the real axis between the
saddle points. At the saddle points it veers off at $60^\circ$ angles, curves,
and asymptotes at $-\pi/4$ and at $-3\pi/4$.}
\label{F2}
\end{figure}

In the correlated limit
the critical point is determined by requiring that $\cL''(r)=0$ in addition to
$\cL'(r)=0$. The critical value of the coupling constant is $g_{\rm crit}=1/4$.
At this value of $g$ there is a coalescence of the two uncorrelated saddle
points at $r_{\rm crit}=\pm\sqrt{2}$, as shown in Fig.~\ref{F2}. The steepest
curve is shown in Fig.~\ref{F2} and the contribution to the integral in the
double-scaling limit comes from {\it four} infinitesimal regions at the saddle
points, two horizontal and two at $60^\circ$ angles. We must now evaluate four
{\it convergent} integrals of the form $\int_0^L dr\,e^{ar^2+br^3}$, where $L=
\infty$ or $L=\infty e^{-i\pi/3}$. Each of these integrals separately gives a
combination of Airy functions and a ${}_2F_2$ hypergeometric function. However,
we will now demonstrate that when the four integrals are combined there is a
dramatic simplification and the result is identical to that in (\ref{e18}).

We begin with the integral $\int dr e^{-r^2+r^3/(3\sqrt{\gamma})}$, where
$\gamma=G^3N/2$ and $G=\sqrt{1-4g}$. Note that the sign of $\lambda$ is reversed
in (\ref{e21}) compared with (\ref{e15}). There are four saddle points of cubic
type. This gives rise to two integrals for each saddle point:
$$\int_{-\infty}^0 dr+\int_0^{\infty e^{-i\pi/3}}dr~+~{\rm complex~conjugate}.$$
In the first integral we replace $r\to-r$ and in the second we replace $r\to
re^{-i\pi/3}$. The first integral then becomes $3^{1/3}\gamma^{1/6}\int_0^\infty
e^{fu^2-u^3}$, where $f=-3^{2/3}\gamma^{1/3}$, and the second becomes $3^{1/3}
\gamma^{1/6}e^{-i\pi/3}\int_0^\infty e^{fu^2-u^3}$, where $f=e^{i\pi/3}3^{2/3}
\gamma^{1/3}$. Next, we use the identity
\begin{equation}
\int_0^\infty e^{fu^2-u^3}=2\pi e^{2f^3/27}3^{-4/3}{\rm Bi}\left(3^{-4/3}f^2
\right)+\frac{f}{3}\,{}_2F_2\left(\frac{1}{2},1;\frac{2}{3},\frac{4}{3};
\frac{4f^3}{27}\right),
\label{e22}
\end{equation}
where the generalized hypergeometric function is defined by the Taylor series
$${}_2F_2(a,b;c,d;z)\equiv\frac{\Gamma(c)\Gamma(d)}{\Gamma(a)\Gamma(b)}
\sum_{n=0}^\infty\frac{\Gamma(a+n)\Gamma(b+n)}{n!\Gamma(c+n)\Gamma(d+n)}z^n.$$

For both the first integral and the second integral, $z=-4\gamma/3$. Thus, the
two hypergeometric functions are real and identical, their sum exactly cancels,
and only the ${\rm Bi}$ functions remain. There are four such Airy functions:
$$2\pi e^{-2\gamma/3}3^{-4/3}\left[ 2{\rm Bi}\left(\gamma^{2/3}\right)
-e^{2i\pi/3}{\rm Bi}\left(e^{2i\pi/3}\gamma^{2/3}\right)
-e^{-2i\pi/3}{\rm Bi}\left(e^{-2i\pi/3}\gamma^{2/3}\right)\right].$$
We simplify this expression by using the identity ${\rm Bi}(z) +\omega{\rm Bi}
\left(\omega z\right)+\omega^2{\rm Bi}\left(\omega^2z\right)=0$ \cite{R6}.
The final result for the integration is therefore $2\pi e^{-2\gamma/3}3^{-1/3}
{\rm Bi}\left(\gamma^{2/3}\right)$. This concludes the demonstration.

\section{Final remarks}
\label{s8}
We have shown that the key to constructing and interpreting the double-scaling
limit is continuing in $\veps$ rather than in $g$. In the past the way to deal
with a negative-quartic potential has been to imagine an analytic continuation
from positive $g$ to negative $g$. However, this procedure would give
complex-energy eigenvalues because the path from positive to negative $g$
crosses branch cuts in the coupling-constant plane, which is a multisheeted
Riemann surface \cite{R13}. We have used a different analytic continuation,
$\veps:\,0\to2$, which does not involve the coupling constant. To do so, we have
written the potential in manifestly $\cP\cT$-symmetric form. The complex
potentials for all $\veps\geq0$ give real partition functions and at $\veps=2$
we have made contact with the quartic theory obtained in the conventional
double-scaling limit.

\vspace{0.5cm}
\footnotesize
\noindent
CMB was supported by the U.K.~Leverhulme Foundation and the
U.S.~Department of Energy.

\end{document}